\documentclass[mypaper,7pt,twoside]{CoAst}
\usepackage{epsf,graphicx,fancyhdr}
\renewcommand{\chaptermark}[1]{\markboth{#1}{}}

\newcommand{\kopf}{\small\itshape Comm. in Asteroseismology \\ Contribution to the Proceedings of the 38$^{th}$\,LIAC\,/\,HELAS-ESTA\,/\,BAG, 2008
}

\newcommand{\Authors}[1]{\begin{center}\normalsize\bf\sf #1 \end{center}}

\renewcommand{\author}[1]{\begin{center}\normalsize\bf\sf #1 \end{center}}
\newcommand{\Address}[1]{\begin{center}\small\sf #1 \end{center}}

\DeclareGraphicsExtensions{.eps,.jpg}

\renewenvironment{abstract}{\section*{Abstract}\normalsize\sf}{}
\newcommand{\References}[1]{\begin{flushleft}{\large References\\}\vspace*{2mm}\small #1 \end{flushleft}}

\newcommand{\chapterCoAst}[2]{\chapter[\sf\normalsize #1\\ \footnotesize \hspace*{5mm}by #2 \sf\normalsize][]{#1\\}\rhead[\fancyplain{}{\sf\footnotesize \center{#1}}]{\fancyplain{}{\sffamily\thepage}}\lhead[\fancyplain{\kopf}{\sffamily\thepage}]{\fancyplain{\kopf}{\sf\footnotesize \center{#2}}}}

\newcommand{\figureCoAst}[5]{\begin{figure}[#4]
\centering
\includegraphics*[#5]{#1}
\caption{#2}
\label{#3}
\end{figure}}

\renewcommand{\chaptername}{}

\newcommand{\acknowledgments}[1]{\vspace*{5mm}\noindent  \textbf{Acknowledgments.} #1}

\def\rfr{\smallskip\par\noindent
        \hangindent=7truemm
        \hangafter=1}

\begin{document}
\sf

\chapterCoAst{Modeling Massive Stars with Rotation: the Case
of Nitrogen Enrichments}
{A. Maeder, G. Meynet, S. Ekstr\"{o}m, C. Georgy} 
\Authors{A. Maeder, G. Meynet, S. Ekstr\"{o}m, C. Georgy} 
\Address{
Observatory of the Geneva University \\
}

\noindent
\begin{abstract}  Recently, the concept of rotational mixing has been 
challenged by some authors (e.g. \textit{Hunter et al. 2008}). We show
that the excess N/H is a multivariate function $f(M, \, age,\, v \sin i,\, multiplicity, \,Z)$. To find a correlation of a multivariate function with some parameter, it is evidently necessary to limit as much as possible the range of the other involved parameters. When this is done, the concept of rotational mixing is supported by the observations. We also show that the sample data are not free from 
several biases. A fraction  of  $\sim20$ \%
of the stars may escape to the relation as a result of binary evolution.
\end{abstract}


\section*{Introduction}

In the late 70', the reality of  mass loss in massive stars was  debated 
as well as its effect on the evolution. When this became accepted at last 
\textit{(Chiosi \& Maeder 1986)}, the next question was whether  
the large differences in the populations of massive stars (for example the WR/O  and WN/WC star numbers)
were due to differences in mass loss with metallicity $Z$. Today, the  debate concerns the reality of rotational mixing and its effects on massive star evolution. History never reproduces itself similarly, nevertheless there is a great parallelism in these debates, which  are normal steps in the progress  of   knowledge and  finally lead to a better understanding of stellar physics and evolution.

 There is an  impressive list of consequences of stellar rotation \textit(Maeder \& Meynet 2000), many of which are supported by observations: about the stellar shape,
 the temperature distribution at the surface,  the mass loss and its asymmetries, on the size of the  cores, the tracks in the HR diagram, the lifetimes,
 the surface composition,  the chemical yields,  the ratios of the different 
kinds of massive stars (blue, red supergiants, WR stars), the types of supernovae
and  the remnant masses, etc. 
In this Liege Colloquium, the reality of the
rotational mixing was  disputed by several authors on the basis of new observations, in particular the VLT-Flames survey \textit{(Hunter et al. 2008)}.
Thus, we concentrate here on this problem, firstly by recalling some 
theoretical predictions  concerning surface enrichments and secondly by carefully examining the observations.

\section*{Recall of theoretical predictions concerning rotational mixing}

\begin{center}
\figureCoAst{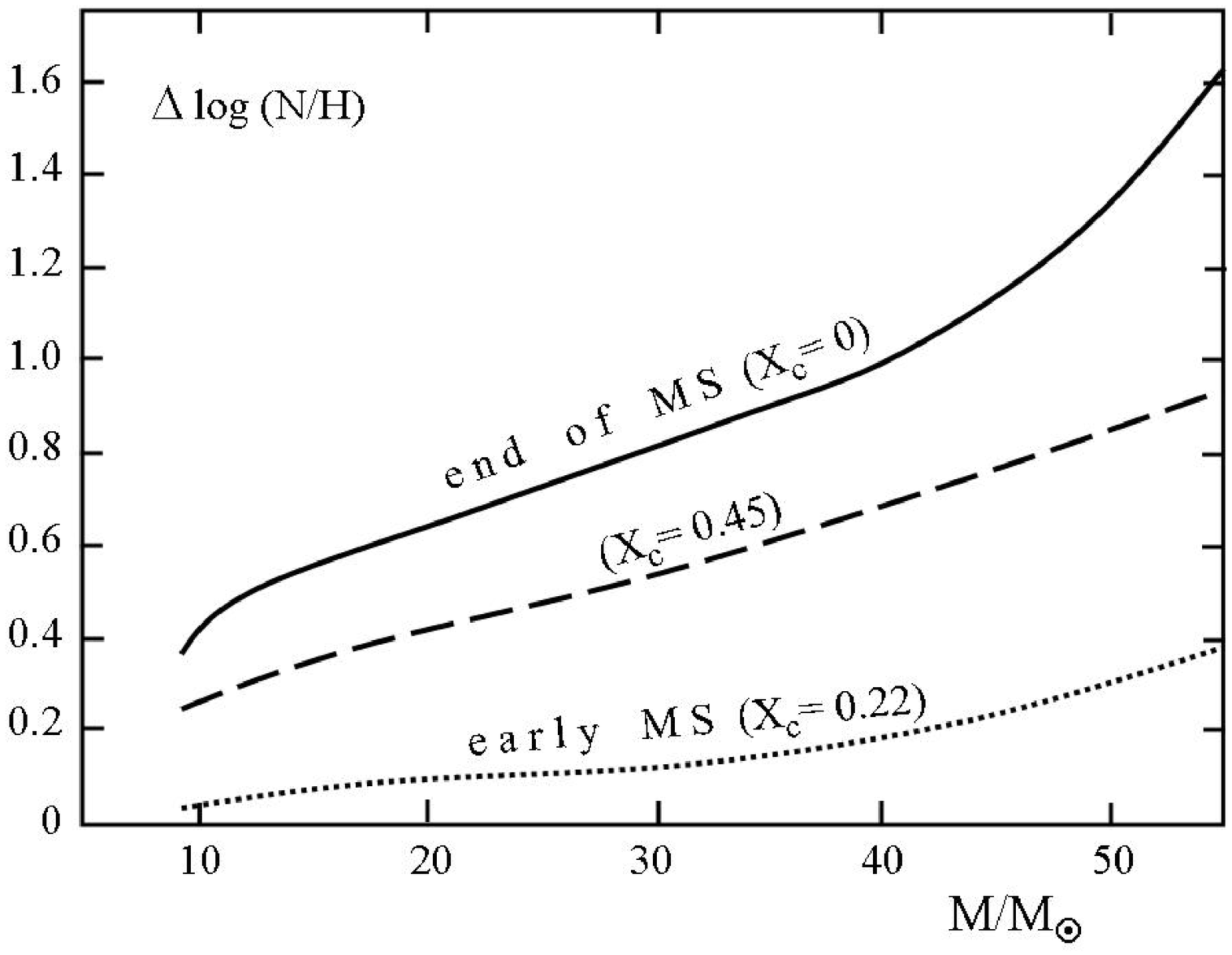}{The differences in log(N/H) as a function of the initial masses at  3 stages during the MS phase for models with $Z=0.02$ with average rotation velocities of 
217, 209, 197, 183, 172, 168 km s$^{-1}$ for respectively 12, 15, 20, 25 40 and 60
M$_{\odot}$. These  3 stages are indicated by the value of the central H--content $X_{\mathrm{c}}$.}{NHmass}{t}{clip,angle=0,height=65mm,width=90mm}
\end{center}

In the mass range of $\sim$10 to 20 M$_{\odot}$  considered here, mass loss has a limited importance during the Main Sequence (MS) phase. The changes of
abundances are expected to be  mainly due to rotational mixing.
The main effect producing element mixing  is the diffusion by shear turbulence,  
which itself results from  the internal $\Omega$ gradients built during evolution.
To a smaller extent, meridional circulation  makes some transport, however mainly of angular momentum. Mixing brings to  the  surface the products of  CNO burning: mainly 
$^{14}$N  and $^{13}$C enrichments, $^{12}$C is depleted with limited $^4$He enrichment and $^{16}$O depletion.

\figureCoAst{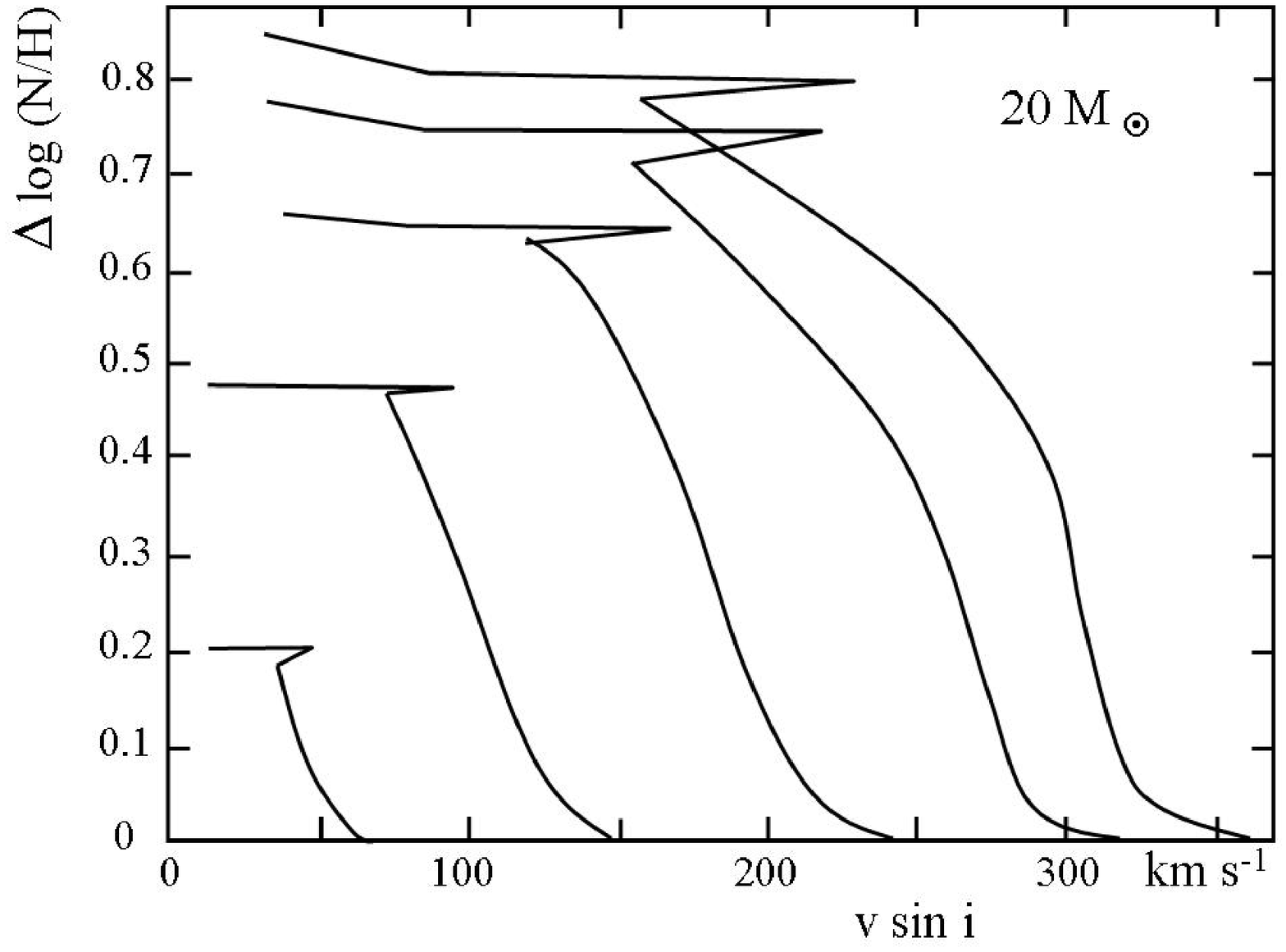}{The evolution of the differences in log(N/H) during the 
MS phase as function of the actual rotation velocities ($\sin i$ being equal to 1 here) for models of 20 M$_{\odot}$ with $Z=0.02$ and  different initial velocities.}{NHvsini}{th}{clip,angle=0,height=65mm,width=95mm}

Fig.~\ref{NHmass} shows the predicted  variations  of  $\log(N/H)$ during MS evolution as a function of the initial masses
\textit{(Meynet \& Maeder 2000)}. (N/H) is here the abundance ratio of N and H in numbers (the relative differences in mass and number are the same).
 Without rotational mixing, there would be no enrichment until the red supergiant stage. Rotation produces an  increase (depending on velocity $v$)  of   N/H 
during  the MS phase. The N excesses also depend on the ages $t$. The increase is modest during the first third of the MS phase, because the elements need some time to reach the surface, then it is more rapid.  The  N enrichments are larger 
for larger  masses $M$. Thus we see that the N excesses are multivariate functions 
\begin{eqnarray}
\Delta \log(N/H) =  f(M, \, t,\, v \sin i,\, multiplicity, \,Z) \;.
\label{f1}
\end{eqnarray} 

\noindent
Models with lower initial metallicities $Z$ have higher 
N enrichments for given $M$ and $v$ (\textit{Maeder \& Meynet 2001;
Meynet et al. 2006}). The excesses become  very strong at metallicities $Z$ as low as $10^{-8}$. 

Fig.~\ref{NHvsini} shows, for  given $M$ and $Z$, the evolution of the N excesses  with age for different initial velocities. This figure  shows that  there can be no single relation between $v \sin i$ and the N/H excesses for a mixture of ages. The same is true for a mixture of different masses. 

Binarity may also affect the N and He enrichments due to tidal
mixing and mass transfer. 
A binary star with  low rotation may have a high N/H due to tidal mixing or due to the transfer of the enriched envelope of a  red giant. At the opposite, a binary star
may also have a high $v \sin i$ and no N/H excess, in  the case of the accretion of an unevolved envelope bringing a lot of angular momentum. A nice illustration
has been given  \textit{(Martin, this meeting)}.

\section*{Study  of the observed N/H excesses}

\subsection*{The $M$, age and  $Z$ dependences}

Let us start by examining the mass, age and $Z$ dependences of the N/H excesses.
The  data for different groups of stars with different $Z$ are summarized in Table 
\ref{tblabindo}. In the Galaxy ($Z\approx 0.02$), the main recent data sources 
\textit{(Herrero 2003; Venn \& Przybilla 2003; Lyubimkov et al. 2004; Huang \& Gies 2006; Trundle et al. 2007)}
 support  significant  excesses of He or of (N/H).  
 In the lowest mass range
studied (6.6--8.2 M$_{\odot}$), small excesses of He/H are still present  
\textit{(Lyubimkov et al. 2004)}. 
In the LMC ($Z\approx 0.008$), the excesses are larger \textit{(Hunter et al. 2007;
Trundle et al. 2007)}.  
In the SMC ($Z\approx 0.004$), still much larger N excesses are observed  
\textit{(Venn, Przybilla 2003; Heap \& Lanz 2006; Trundle et al. 2007; Hunter et al. 2007)}.

\begin{table}[!h]  
 \caption{The largest $\Delta \log (N/H)$ values observed  for different types of stars 
 in the Galaxy, LMC and SMC (differences in dex with respect to the local values   in the considered galaxy).
 The average is equal to about the half of the indicated values.} \label{tblabindo}
\begin{center}\scriptsize
\begin{tabular}{lccc}
                &                     &                &               \\
Types of stars     &$\Delta \log (N/H)$ Galaxy  &$\Delta \log (N/H)$ LMC&$\Delta \log (N/H)$ SMC\\
                &                     &                &               \\
\hline 
                &                     &                      &               \\
  O stars                            & 0.8 - 1.0            &   --           &  1.5 - 1.7     \\
  B--dwarfs $M<20$ M$_{\odot}$         & 0.5                 & 0.7 - 0.9      &  1.1            \\
B giants, supg. $M<20$ M$_{\odot}$  & --                  & 1.1 - 1.2      &  1.5            \\
B giants, supg. $M>20$ M$_{\odot}$  & 0.5 - 0.7           & 1.3            &  1.9            \\
                &                     &                      &               \\
 \hline    
\end{tabular}
\end{center}
\end{table}

\noindent
These  data show the following facts, consistent with theoretical predictions:
\begin{itemize}
\item On the average, the N enrichments are larger for larger masses.
\item  The N enrichments are larger at lower $Z$.
\item  The  He and N enrichments increase with the distance to the ZAMS \textit{(Huang \& Geiss 2006)}. They are even larger in the giant and supergiant stages \textit{(Venn \& Przybilla 2003)}. This property is also well observable for example in
N11 (Fig.~34 of \textit{Hunter et al. 2007}) and in  NGC 2004 (Fig.~2 by \textit{Trundle et al. 2007}).
\end{itemize}
 
 \subsection*{The $v \sin i$ dependence of the N/H excesses}
 
 Several correlations of the N or He excesses with the observed $v \sin i$ have been performed. \textit{Huang and Gies (2006)} and \textit{Lyubimkov et al. (2004)} find a correlation of the He excesses
with $ v \sin i$ for B stars in the upper part of the MS band in agreement with model predictions. 

In other comparisons \textit{(Hunter et al. 2008; Langer, this meeting; Brott, this meeting)}, the authors conclude that ``the observation \ldots challenges the concept of rotational mixing''. They claim that ``two groups of core hydrogen burning stars
\ldots stand out as being in conflict with the evolutionary models''. Group I
contains rapid rotators whith  little chemical mixing, while  Group II consists of low rotators with large N enrichments.
We clearly disagree with the conclusions of \textit{Hunter et al. (2008)}, which mainly result from the fact that, instead of  Eqn. (\ref{f1}), their analysis  implicitly assumes  that 
\begin{eqnarray}
\Delta \log(N/H) =  f(v \sin i) \;.
\label{f2}
\end{eqnarray}
\noindent
We note the following points: 
\begin{itemize}
\item Their  sample  contains a mixture stars in the mass
interval of 10 to 30 M$_{\odot}$. Fig.~\ref{NHmass} shows that over this mass interval, the N/H excesses vary as much as by a factor of two for a given rotation velocity.
\item The sample by Hunter et al. consists of stars in extended regions around the LMC clusters N11 and NGC 2004. As stated by the authors, their sample also contains field
stars, which do not necessarily have the same age or degree of evolution as the cluster stars. Thus, \emph{large differences of N/H are possible for given $M$ and $v \sin i$}.
Also,
the two clusters do not have the same ages, N11 being younger than NGC 2004, so that the  stars near the turnoff of N11 have a mass of about 20 M$_{\odot}$, while this is about 14 M$_{\odot}$ for NGC 2004 according to the HR diagrams by 
respectively \textit{Hunter et al. (2007)} and \textit{Trundle et al. (2007)}.
\item The completeness of the binary search is unknown. 
\end{itemize}
\vspace*{-8mm}
 
\begin{center}
\figureCoAst{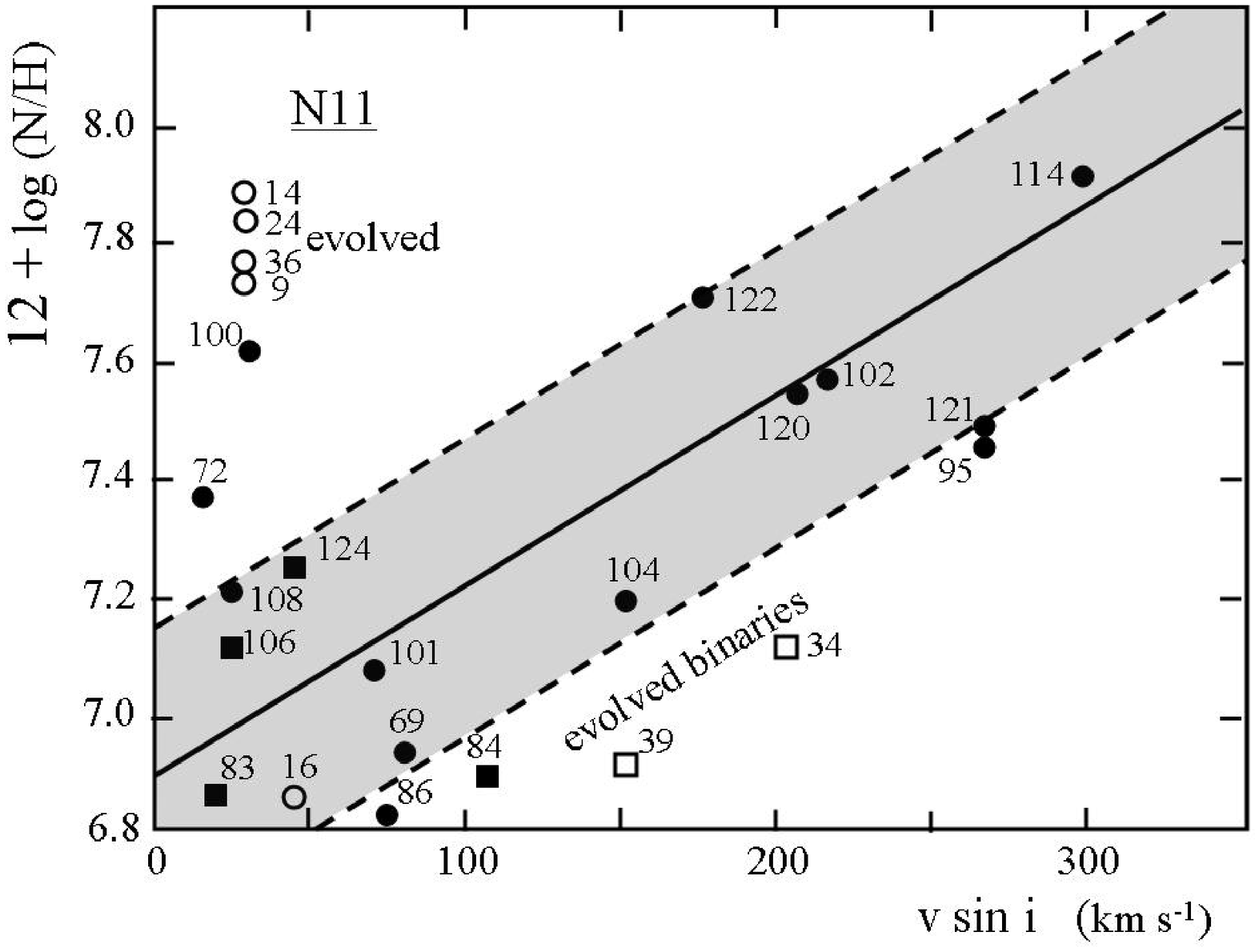}{The N abundance (in a scale where $\log H=12.0$) 
as a function of $v \sin i$ for the MS stars (black dots) in N11 with masses between
14 and 20 M$_{\odot}$ according to \textit{Hunter (2008)}. The binaries are shown by a square. The evolved stars
in a band of 0.1 dex in $\log T_{\mathrm{eff}}$ beyond the end of the MS are shown with open symbols.
The gray band indicates uncertainties of $\pm 0.25$ dex.} {NHN11}{t}{clip,angle=0,height=65mm,width=90mm}
\end{center}

To limit the severe effects of mass and age differences, we consider separately the two clusters. In N11, we limit the sample to the stars in the mass range 14 to 20 M$_{\odot}$ on the basis of  the data provided by \textit{Hunter (2008)} and in the formal MS band  as given by Fig.~34 from \textit{Hunter et al. (2007)}. In  NGC 2004, we take the mass interval 13 to 16 M$_{\odot}$ (same source)
and in the formal MS band from Fig.~2 by \textit{Trundle et al. (2007)}. 
Figs. \ref{NHN11} and \ref{NHNGC2004} show the results. Ideally these mass intervals should even be smaller.

\figureCoAst{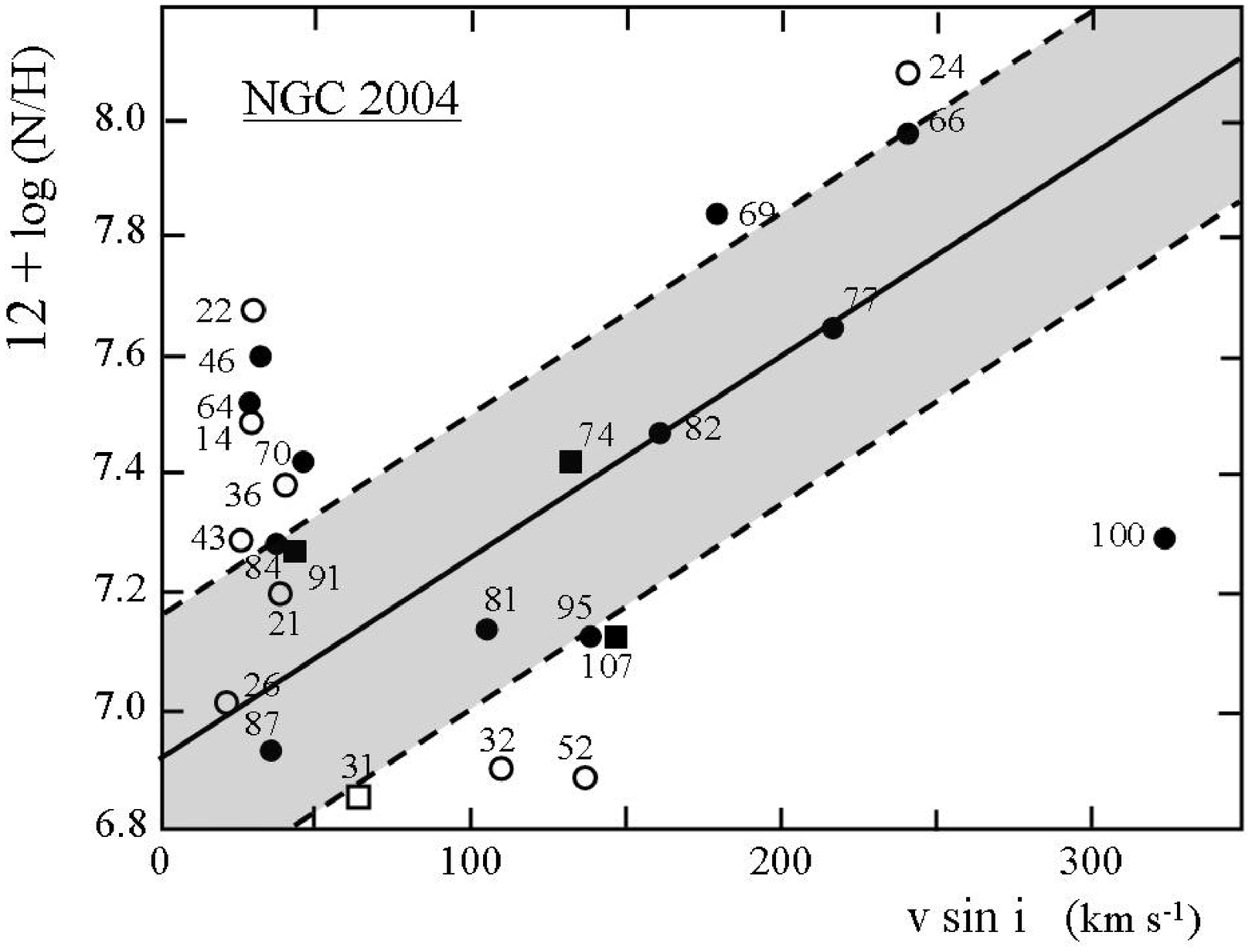}{The N abundance 
as a function of $v \sin i$ for the MS stars in NGC 2004 with masses between
13 and 16 M$_{\odot}$. Same remarks as for Fig.~\ref{NHN11}.}{NHNGC2004}{t}{clip,angle=0,height=65mm,width=90mm}

For N11, we see that Group I (stars with high $v \sin i$ and low N/H) has essentially
disappeared. There remain  only 2 evolved binary stars, which is  consistent with
some scenario of binary evolution. We suspect that Group I was also largely 
formed by stars of smaller ages and/or lower mass stars, where for a given $v \sin i$ the N excesses are smaller.
A support to the latter possibility comes from the fact that the average mass of the stars in the region of Group I
 is 12.8 M$_{\odot}$ (for $v \sin i > 180 $ km s$^{-1}$ and  $12+\log(N/H)< 7.30$)  and 17.1 M$_{\odot}$ for stars 
with $v \sin i > 180 $ km s$^{-1}$ and  $12+\log(N/H)> 7.30$. Group II is mainly made from evolved stars (open circles), which explains the low velocities and high N abundances.
The two remaining stars in this group can easily be stars with a small $\sin  i$, especially more
than the sample data by \textit{Hunter et al. (2008)} is largely biased toward low rotators. The ratio of star numbers  with $v \sin i \geq 250$ km s$^{-1}$ to those with $v \sin i \leq 100$ km s$^{-1}$is 0.14, while in the clusters studied by 
\textit{Huang \& Gies (2006)}, this ratio amounts to 0.40! Thus, the sample by \textit{Hunter et al.} contains a large excess of slow rotators.
We conclude
that the bulk of stars in N11 shows a relation of the excess of N/H depending on 
$v \sin i$ (the mean square root of the data for the MS band stars is 0.23 dex from the data by \textit{Hunter (2008)}, the scatter in $v \sin i$ is not given). The  amplitude of  the (N/H) is about 0.6 dex for velocities of  200 km s$^{-1}$,  slightly   higher than at the corresponding mass in  Fig.~\ref{NHmass}  for $Z=0.02$.

For NGC 2004, the results are essentially similar. For most stars there is a relation between the excess of N/H and $v \sin i$. For Group I, there remains
only star nb. 100 (which is not much for a group). This star is interesting.
Its mass is 13 M$_{\odot}$, $v \sin i$ is 323 km s$^{-1}$, the highest of the whole sample. In reality, the velocity is  still higher because the authors do not account for gravity darkening  (Fig.~4 of \textit{Hunter al. (2008)}
tends to support this remark). This star might be a re--accelerated binary or
simply a younger star in the field.
 Another possibility (which we favour) is that its parameters have been incorrectly appreciated due to the
extreme rotation. In this respect, a $\log g$ vs. $\log T_{\mathrm{eff}}$ diagram,
e.g. Fig.~16 by \textit{Meynet \& Maeder (2000)}, shows that a too high mass is assigned to a fast rotating star if in the $\log g$ vs. $\log T_{\mathrm{eff}}$ diagram its mass is determined from non rotating models. Thus, the mass of star nb. 100 could be lower than 13 M$_{\odot}$ which is the lower bound of our sample. To know whether this is what occurred for this star, the whole reduction process should be redone.
For Group II, a large fraction consists as for N11 of evolved stars. Again for
NGC 2004, the N enrichments increase with $v \sin i$ in agreement with theory.

We also note that  the  data used in the analysis by \textit{Hunter et al. 
(2008)} are subject to several biases. Firstly, the sample contains no Be stars, while their number fraction is about 15 to 20 \% in the LMC. This contributes to bias the sample toward low velocities.
A second source of bias is that the $v \sin i$ determinations are based on models assuming that the stars are uniformly bright, with no account  given to  gravity darkening (now an observed effect). Thirdly, in the values of gravity used to estimate the masses, no account is given to  
the gravity change due to rotation.  The effects of evolution and rotation should be disentangled before any  mass is assigned.

\section*{Conclusions}
We conclude quite logically that to find a correlation for a multivariate function like N/H with some parameter like $v \sin i$, it is necessary to limit as much as possible the range of the other involved parameters. Otherwise, the conclusions may
be erroneous. We note that data samples limited in mass and ages support a N enrichment depending on rotational velocities. Stars beyond the end of the MS phase
do not obey to  such a relation, because their velocities 
converge toward low values (see Fig.~12 by \textit{Meynet and Maeder 2000)}. 
A fraction, which we estimate to be   $\sim20$ \%
of the stars, may escape from the relation as a result of binary evolution, either by tidal mixing or mass transfer.

\acknowledgments{We thank Dr. I. Hunter for having provided the observed data used in  
this study.}

\References{
\rfr Chiosi, C., Maeder, A., 1986, Ann. Rev. Astron. Astrophys. 24, 329
\rfr Heap, S.R., Lanz, T.,  Hubeny, I. 2006, ApJ {638}, 409
\rfr   Herrero, A. 2003, in \textit{CNO in the Universe}, Eds. C. Charbonnel, D. Schaerer, G. Meynet,
ASP Conf. Ser. 304, 10  
\rfr Huang, W.,  Gies, D.R. 2006, ApJ 648, 580 \& 591
\rfr Hunter, I. 2008, private communication
\rfr Hunter,I.,  Dufton, P.L., Smartt, S. et al. 2007,  A\&A {466},
277 
\rfr Hunter, I., Brott, I., Lennon, D.J. et al. 2008, ApJ, 676, L29 
\rfr Lyubimkov L.S.,  Rostopochin, S.I.,  Lambert, D.L. 2004, MNRAS {351}, 745 
\rfr Maeder, A., Meynet, G. 2000, Ann. Rev. Astron. Astrophys., 38, 143
\rfr Maeder, A., Meynet, G. 2001, A\&A, 373, 555
\rfr Meynet, G., Maeder, A. 2000, A\&A, 361, 101 
\rfr Meynet, G., Ekstr\"{o}m, Maeder, A. 2006, A\&A, 447, 623
\rfr Venn, K.A.,   Przybilla, N. 2003,   in \textit{CNO in the Universe}, eds. C. Charbonnel, D. Schaerer, G. Meynet, ASP Conf. Ser. 304, 20 
}

\end{document}